# Assessing the Usages of LMS at KAU and Proposing "FORCE" Strategy for the Diffusion

**Rayed AlGhamdi**[1] and **Adel Bahadad**[2]

[1] *Information Technology Department, and* [2] *Information System Department, Faculty of Computing and Information Technology, King Abdulaziz University, Jeddah, Saudi Arabia*

raalghamdi8@kau.edu.sa

*Abstract*. Since the beginning of the Saudi Arabian academic year 1435 (Sept 2014), the web-based learning management system Blackboard has been introduced and made available to all instructors and students for all courses at King Abdulaziz University (KAU). The current study takes place to assess the current usages of the Blackboard usages at KAU. The data collected from the 923 students of the foundation year which represent about one-third of the total number of the male students for the academic 2016/2017. Based on statistical evidence gained from the students' responses to the survey questions, 78% of the students are inactive users of the Blackboard. The study follows up with interviewing five instructors who teach first-year students in order to seek explanations of the Blackboard low usages by the students. The outcomes point significant processes at an individual level and as well as an organizational level. The Diffusion of Innovation (DOI) was used to study the case because it is believed to be the best explain such adoption of innovation at individual and organizational levels. Based on the current outcomes and the author's experience in teaching a computer course using Blackboard, a strategy called 'FORCE' is proposed for the diffusion process.

## 1. Introduction

In the last two decades, higher education institutions around the world have been accelerating the use of technology in education. The use of technology in education differs from university to another; however, in general, the purpose is to reach a wider audience and to serve multiple purposes of teaching and learning processes [1]. A popular computerized system called Learning Management System (LMS) has been widely adopted by universities. LMS allows instructors to provide and manage courses and training programs in organized forms that facilitate for the students to access through the Internet and interact based on every individual pace to achieve learning objectives [2]. The main target of such systems is on online learning delivery; however, they are used widely as a platform for traditional education and some hybrid forms in between online and traditional including blended learning, active learning, collaborative learning and flipped classrooms [3].

There are various products of LMS available on the market. For example, the top three LMSs products in the USA are Blackboard, Moodle, and Canvas with the market share 33%, 19%, and 17% respectively [4]. Blackboard has had attained a significant global market share. In Saudi Arabia, for example, the ministry of higher education has signed a contract with the Blackboard Inc. to introduce the system to all the 28 government universities [5, 6]. For few years before using the Blackboard, KAU used the Moodle system only for distance learning students [7].

Regardless the name of a system used to manage the teaching and learning process, introducing such a system into a social environment involves various aspects that should be considered. What about if you introduce a new system to an organization and people resist using it or the usages do not fulfill the expectations? The human element is not a component to match with a system to basically switch on/off, it is much more complex.

## 2. Related Studies

The ongoing technology evolution in education places additional demands upon educational institutes to keep up with this pace; and provide their instructors and students with the necessary resources and training to ensure successful acceptance and use. Al-Malki, AbdulKarim & Alallah [8] conducted a study to measure the KAU distance-learning teachers and students' satisfaction using the Blackboard. The study indicates a relatively high satisfaction using Blackboard in the distance educational process. Distance-learning is completely online at KAU and the main mean to deliver the courses is Blackboard. Using Blackboard in distance-learning at KAU is compulsory for both teachers delivering lectures and provide materials, and as well students to gain learning materials. A question might be asked here: what if are there options other than Blackboard? This is applicable to traditional education (face-to-face classes). The Blackboard is available for all face-to-face courses at KAU. What is the percentage of using the Blackboard system to enhance traditional education and to what extent is it effective? Based on our best practice searching for literature that covers these issues, nothing has been found. Most of the literature paid high emphasis seeking users' perceptions towards using LMS in education. For example, Alsaied [9] conducted a study to figure out the teachers' perceptions toward using Blackboard at KAU English Language Institute. The study surveyed 40 teachers to come up with the result that the participants have positive trend to use the Blackboard in the teaching process. In addition, Binyamin, Rutter & Smith [10] investigated the factors that Influence the LMS usages from the KAU students' perspectives. They confirmed that the prior experiences, satisfaction, social influence, computer self-efficacy and teacher role are significant factors that influence students to accept/reject using a learning management system such as the Blackboard system.

At a global level, a great deal of the literature has investigated the acceptance, adoption and use of various educational technologies. Brown [2] provided a solid review of the literature on instructors' adoption and use of online technology in traditional education. He classified the influences into external and internal. The external influences are all around interactions with technology, academic workload, institutional environment, and interactions with students; whereas the internal influences relate to instructors' attitudes and beliefs, and experience [2].

Most of the reviewed literature paid high emphasis on the instructors adopting new technology [4, 11, 12, 13, 14 & 15]. However, introducing a new technology to an organization involved two levels: organizational and individuals. Therefore, studying the process of adoption should consider all the relevant entities.

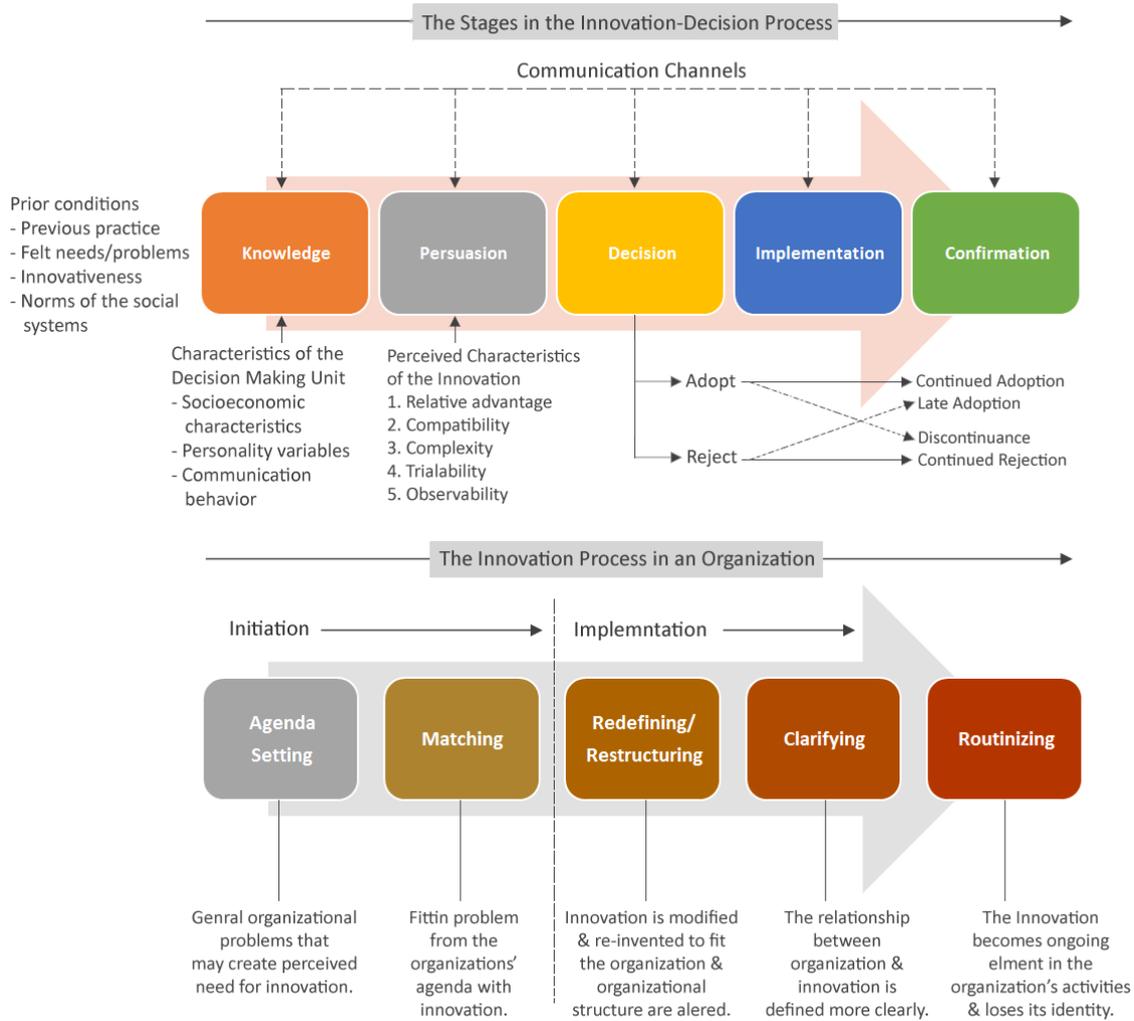

**Fig. 1.** The processes of Innovation Adoption at individual level and Organizational level. [16, p. 170 & p. 421].

We believe considering the Diffusion of Innovation (DOI) theory to study the process of LMS adoption and diffusion at KAU is worthy. DOI provides an inclusive view of the process of innovation decision-making and undertakes the explanation how an idea, practice, or object that is perceived as new by an individual, or another unit of adoption, is spread [16 & 17]. In our case, the implementation of LMS at KAU involves individuals and organizational entities. DOI is effective in studying this case because it studies the phenomenon of introducing a new technology at two levels: individuals and organizations; see Fig. 1.

Rogers [16] defines diffusion as "the process during which an innovation is communicated through certain channels over time among members of a social system". These elements are (1) the innovation, (2) communication channels, (3) time, and (4) social systems. An innovation is not necessary to be totally an invention but rather an idea, practice, or object that is 'perceived' by an adoption entity as new [16]. At the individual level, the DOI identifies five stages that an individual goes through to reach a decision whether to adopt or reject a new technology. The first stage is the knowledge about the new technology. To what extent an individual has enough knowledge to be capable to understand what this new technology can be used for. An ability to obtain knowledge differs from a person to another. These differences are based on people personalities, their needs, social systems norms, communication behaviors etc. Once an individual or adoption unit has minimum knowledge, they move to the persuasion level. To what extent they can be convinced that the adoption of such a technology serves their needs. The decision at this stage is influenced by how an adoption unit perceives the use of a new technology. Will the use of this new technology provide advantages, will be compatible with the current system/uses, is it easy to use, can be tried and observed? These five attributes relate to the newly introduced technology and play a significant role to convince the adoption decision. Once the user is convinced, it is most likely the adoption process will be successful.

At an organizational level, DOI identifies other five stages of the process innovation. These stages are divided into two major phases: initiation and implementation. The initiation phase involves two stages: agenda setting and matching; whereas the implementation phase involves three stages: redefining/restructuring, clarifying, and routinizing. It is not enough for an organization to introduce a new technology and directs its entities to start using it. When an organization felt there is a need to use a technology, it is not basically the decision that needs to be taken and that's it. The process of an innovation adoption at the organizational level is more complex. It requires changes in the organization itself. Innovativeness of an organization is related to three independent variables: management characteristics and attitude toward change, internal characteristics of organizational structure and external characteristics of the organization [16], see Fig. 2.

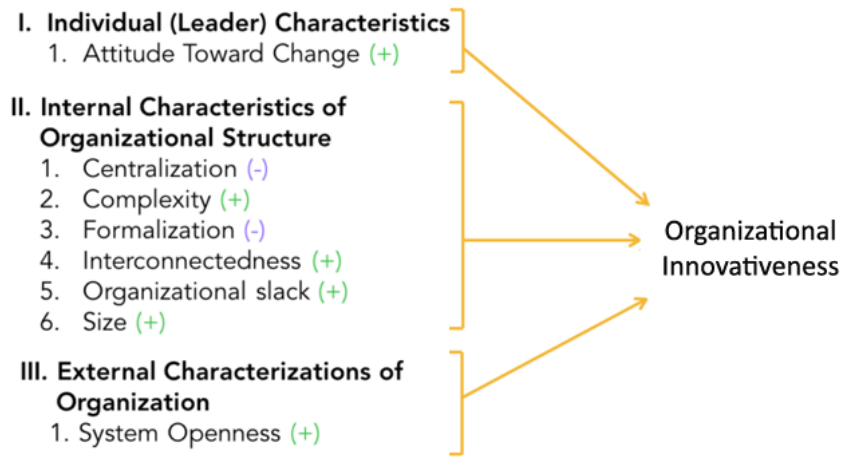

**Fig. 2. Variables determining organizational innovativeness, [16, p. 411].**

We believe using the DOI in the current study is the most relevant. DOI model is the best conceptual model to explain the adoption process by individual and organizations over time [18 & 19]. It has been widely used to explain the adoption of innovations, especially those involving technology [18, 20 & 21]. It assists examining different aspects, innovations characteristics, decision type, communication channel, social system nature, and efforts made by change agents, which influence decision-making.

## 3. Methodology

Since the study seeks explanations on the question why the diffusion process of LMS adoption is slow, the explanatory mixed methods design was adopted. The mixed methods design relies on collecting quantitative data first, analyzing these data, and then collecting qualitative data to explain the quantitative results [22]. A short questionnaire was designed to collect empirical data from the targeted sample, KAU foundation-year students. Interviews with faculty members were used to collect qualitative data for obtaining explanations.

The first phase of data collections involved collecting data from the KAU foundation/first-year students. A short survey was designed to collect data about the Blackboard usages. The survey starts with a straight forward question: do you use Blackboard in your courses? Three answers available: "Yes, No, I do not know Blackboard". For the participants opt 'Yes', they were requested to answer three more questions: how regularly do they use (every class, once a week, once a month, or once during the semester), for what purposes do they use (read information about courses, uploads and download files, conduct exams, use communication tools, use discussion board, and other usages), and for which courses are used (a list of all the foundation-year courses was provided [23]).

The survey was distributed in person in May 2017 during the final exam of computer skills course (CPIT-100). The timing of distribution was chosen carefully where all students should gather, in groups between 8 am to 1 pm, in a hall and wait for about 15 minutes before entering classrooms to conduct the exam. A thousand survey forms were distributed. A total of 923 completed forms were returned, giving a response rate of around 92%; whereas the reset 8% of the forms were either incomplete or unreturned.

After analyzing the quantitative data collected in the first phase, interviews were conducted with five instructors who teach foundation-year students. The selection of participants was selective in order to cover active and inactive instructors using the Blackboard in their teaching process. The purpose of the interviews was to seek explanations and reflections on the students' responses to the survey. The interview was unstructured. Its process began with presenting the students' survey results to the interviewee and start having conversations. Each interview session was recorded and took time 15-30 minutes.

## 4. Results and Discussion

### a. Students questionnaire

The statistical data collected from the foundation-year students demonstrate that the vast majority (78%) of the students were classified inactive users of the Blackboard; see Fig. 3. If a student never accesses or only accesses the Blackboard system no more than once during the semester, he is considered inactive user. By contrast among the 202 active users (22%), a tiny number (23 users) that used the system more than once weekly, 89 accessed once a week, and the remaining (90 users) mentioned that they accessed once a month. The purposes of accessing the Blackboard by the users that were classified active are as follows; 189 read information about courses, 108 uploads and download files, 74 conduct exams, 55 use communication tools, 20 use discussion board, and 5 mentioned other usages.

The statistics demonstrate clearly that the computer skills course (CPIT-100) is the most accessed course on the Blackboard by the active users with a rate of 82%, see Fig. 4. The other courses are unattractive to be accessed by active users on Blackboard with a rate less than 30%. Either there was no content for those courses, the content was poor, or the content seemed for extra activities. It seems the reason behind the highest access by active users for the CPIT 100 course is because the course has a chapter on how to use the Blackboard [24]. However, when comparing the 166 users who accessed the course to the total number of the sample, it is only 18%! This percentage clearly indicates that the content on Blackboard of the CPIT-100 course is almost similar to the other courses. Despite the course has a whole chapter about Blackboard, it was not a requirement for the course to be active using the system and show real usages.

An interesting finding relates to the English courses. Despite there are four English courses (ELI 101, ELI 102, ELI 103, ELI 104) are taught for the KAU foundation-year students, the current statistic shows low indicator using Blackboard for English courses, see Fig. 4. However, the literature indicates that the English Language Institute teachers at KAU have positive trend to use the Blackboard in the teaching process [9]. This confirms that the actual use of a system is not relying only on a positive attitude toward using a system.

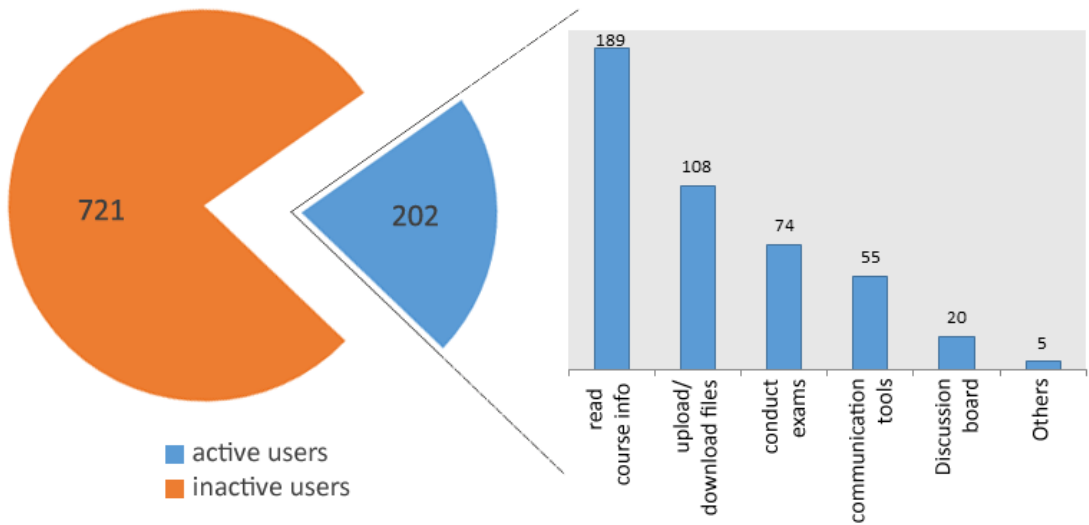

**Fig. 3. Active & inactive users of the Blackboard system & usages of the KAU foundation-year students.**

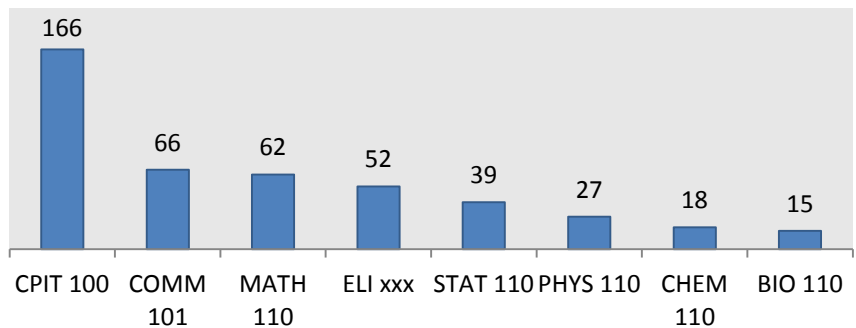

**Fig. 4. The accessed courses on Blackboard by the active users.**

The above statistics prove that the adoption process of the Blackboard system in the KAU foundation-year is not successful yet. Based on the DOI categorization of adopters [16]; see Fig. 5. , it still far from reaching the early majority of adopters. The people involved in an innovation adoption process categorized into five groups: innovators, early adopters, early majority, late majority and so-called laggards. Reaching the diffusion stage is when the early majority complete the adoption [16].

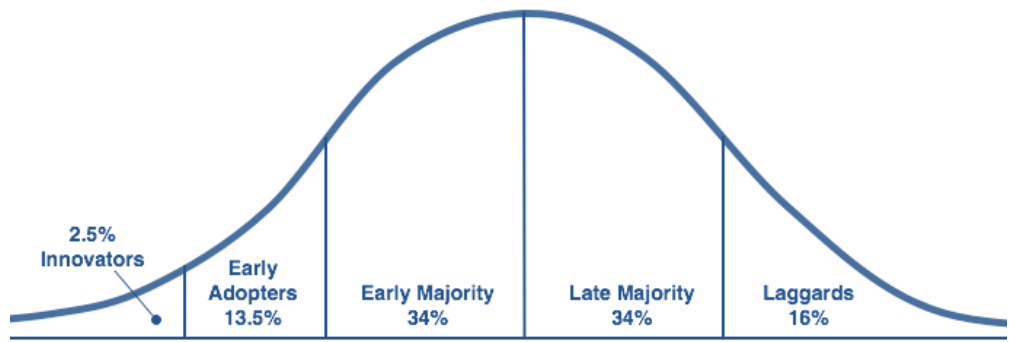

**Fig. 5. Adaptor categorization based on innovativeness [16].**

In this case of not reaching the diffusion stage despite about three years old of the system at KAU, the students cannot be blamed when there is no point to access the course page on the Blackboard system. A part of the story depends on the instructors. Do the instructors provide valuable content for their courses on the system? Is the access to the content on the system a requirement for the course? Is the system being the unique place where the students can access the course material? Answers to more or less of these types of questions assist in clarifying part of the whole image. However, what about if the instructors, similar to the students, see no point in using the Blackboard system?!

*b. Instructors Interviews*

To illustrate the story, interviews were conducted with five instructors who teach courses for the foundation-year students. The selection of the participants was purposive in order to cover instructors use the Blackboard in their teaching process and as well as instructors who do not. Among the five interviewees, one participant fully uses the Blackboard system in all his teaching process (Fahad), the second one use it mostly (Adel), the other two have not used it at all (Naser & Jalal), and the fifth one used it rarely (Saad). In the following paragraphs, the interviewees' discussions are presented. Each of the interviewees was given an unreal name for identification purpose while keeping participants' anonymity. It was interesting to interview an instructor fully uses the Blackboard system in his teaching process to know how it is used. The interview with the instructor (Fahad) gained rich information regarding the Blackboard usages. He started using the system since it has been introduced in 2014. He is considered among the early adopters based on the Rogers' adoptors categorization; see Fig. 5. above. The discussion with Fahad was full of enthusiasm. He taught the computer skills course and was fully self-motivated adopting the Blackboard in his teaching process. He started explaining why he uses this system. He stated that "*the students differ in their skills of the computer, you may find a skilled student compared to a student who even does not know how to start*

*MS Word application! In the normal way of teaching delivering information seems at one pace, mostly in the middle level. A skilled student may find it boring and unskilled student may find it hard to catch up! In pedagogy, this is called the individual differences. Individual differences are hard to be taken into account using the traditional way of teaching especially in teaching or what I prefer to call it training the skills like the case in my course. Blackboard offers options to consider these differences among students. Smart students can complete their tasks quickly and leave even if it within 5 minutes while students who need more attention take their time and get the chances to ask. In my course every class there is an activity. A student is attended if an activity task is completed and uploaded to the Blackboard; otherwise, he is considered absent. I do not have an attendance list! My students are free to come to the class or stay at home. The most important is to complete the tasks on Blackboard. The attendance is calculated based on the completion of the task on Blackboard, no matter where your body is!*".

Interestingly, the second participant (Adel) was influenced by the first one (Fahad). As he stated he was involved in a workshop and private sessions with Fahad to become closer to using full functionalities of Blackboard in his teaching process. Adel commented that "*I was not aware of the Blackboard benefits in the teaching process. I used to use power point slides and display in the classes. I was invited to a workshop on how to use the Blackboard in teaching the computer skills course. Since then I am using it, and I hope I can use full functionalities of this system. The workshop was a starting point and then I discovered that it helps me a lot in my teaching process and save my time.*" The question which might be raised here is that to what extent self-motivation and encouragements could help? Adel replied: "*self-motivation is not enough; it does not work for everyone. The strategy of carrot and stick should be considered by the decision makers... Why do not people use such a system? This is may refer to the culture and habits of people. I remember when I was studying overseas, I used to ask people when I do not know about places, directions… and the common answer is: 'Google it'! This is the culture and habits of people over there. Everything should be online and do not ask before you search on the Internet. Back here in Saudi Arabia, I am trying to reflect the same thing with my students using Blackboard. When they ask about the course syllabus, exam time, assignments, marks distribution… I keep referring them to the course page on Blackboard; no direct answer is given*".

It is clear from Fahad and Adel experiences that they are not limiting their teaching style with what it is so called "*organizational literal rules*" such as accounting students on an attendance list. Apart from management view, Fahad's flexible way combines interaction and attendance. A student must be active on Blackboard to be accounted in the attendance list. It is totally the opposite situation with another participant (Naser) and it seems an obstacle for him to use Blackboard. Naser has negative perceptions towards using Blackboard in his current teaching process because it adds "*extra load*". He said that "*each minute of the lecture is assigned with a part of the content. There is no time for extra activities. If the lecture is 80 minutes, using Blackboard takes some time and the lecture lessons are designed for the 80 minutes, no single minute can be wasted*". It is really surprising to what extent Naser considers the Blackboard system as a waste of time! However, it seems lack of experience creates this negative attitude. Naser indicated that if the Blackboard becomes an essential component of the KAU teaching and learning process, he needs "*assistance on how to use it in practice*". He is concerned about training students as well, "*I need the students to be trained on using it… who is going to train them?!*" It is obvious with Naser's case; he is fixed with the rules. He insisted several times on enforcing the use of Blackboard… "*The management should activate using it in our teaching process… it should be compulsory for all not leaving it an option*". It seems his literal follow of the rules narrows down his thoughts to find ways on how to use the Blackboard with the current rules… "*The course content needs to be developed to be familiar with the use of Blackboard… It should be there some marks to use Blackboard. I cannot change the grading distribution because the distribution is fixed by the department*".

The Naser's case is almost similar to Jalal's case. Jalal has negative perception toward using the Blackboard in his current teaching process due to efforts of extra load that might be needed. He commented: "*There is no need to use it in a face to face education, for e-learning and distance learning yes, it is useful. For a face to face education it consumes lecture time... If the students upload files, it takes time to open, check and mark every single file; this is extra load to my lecture. Normally I check every student work during the lecture time and complete the task within the lecture session.*" Clearly there is a lack of experience with Jalal's case. He judged the usages of Blackboard even before he gets involved! He considered Blackboard mainly as communication tool; he compared it to WhatsApp! "*Most of the students do not use computers at homes, they use their smartphones and I have WhatsApp to communicate with them*", Jalal said. Even he went further in his judgment telling that "*the problem is with the students they will not use it, it is much easier to use their smartphones using any useful application…*" When he confronted with the system benefits and the usages are far more than a communication tool, he acknowledged his lack of knowledge and experience; "*we need more workshops and training sessions for both instructors and students*", Jalal commented.

The fifth participant (Saad) seemed less negativity compared to the previous two cases toward using the Blackboard. Saad said, "*I used Blackboard at the first time I thought it is a requirement for the teaching process at the University*". It looks Saad wants it a top-down decision to use the system. This is not much different from Naser's opinion that "*it is not officially announced this is a requirement for teaching*". It looks he is influenced by the environment surrounding him as he stated "*I know some use Blackboard but most of the instructors do not use it. I do not see many people using it*". In addition, Saad has a similar perception to Naser and Jalal about using the Blackboard in the teaching process because "*it consumes time and needs more efforts when adding this component to my normal teaching process*", Saad said. This might relate to the habit that people used to do and familiar with, fear of exceeding the 'comfort zoon'. "*My students and I are familiar with using e-mail, drop-box; why adding more headaches... Why do I have to use it while the way I used to do in my teaching is work*ing *fine*", Saad commented. Saad insisted many times in his conversation that the Blackboard is not suitable for his course! "*The nature of my course does not require using Blackboard. Using white-board and marker, and some PowerPoint slides to explain the course concepts are enough. Blackboard is not suitable for many courses*" Saad declared. Giving a strong statement or judgment based on poor experience seems odd at an academic level. At least what Saad said is right for his current case/process but not necessary for the nature of the course or other courses. These are interesting and obvious cases that should attract the attention of the decision makers on how to work on the diffusion of the Blackboard system.

### c. *The stages in the innovation-decision process and rate of adoption*

Based on DOI five stages of the innovation-decision process at the individual level [16]; obviously, the first two stages (knowledge and persuasion) play significant roles in the users' decision whether to adopt or reject an innovation. When mapping the study's participants (instructors) to these stages, see Fig. 6. , the negative attitude and lack of experience significantly hinder three participants to adopt the Blackboard system. When working with people like Naser, Jalal and Saad, more efforts are needed on developing their knowledge and enhance their experiences. This is what currently the KAU Deanship of e-learning trying to do through organizing and running different workshops [25]. However, we are not working here with individuals to develop skills for personal uses. The process of teaching and learning involves teachers, students, courses, programs, faculties... and that's why other factors should be considered to work at the organizational level.

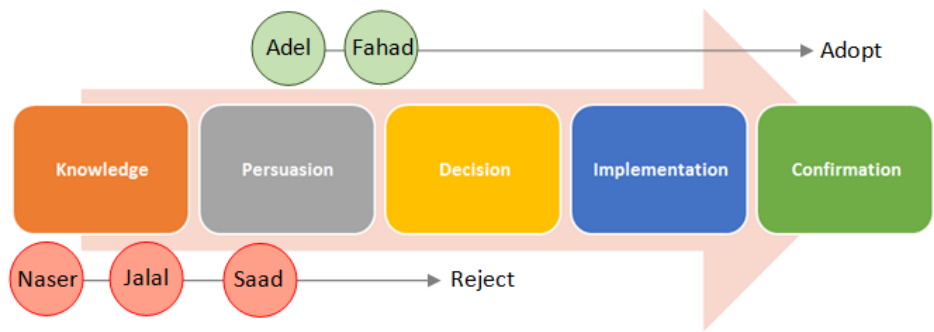

**Fig. 6.  Mapping participants to the five stages of the innovation-decision process at individual level.**

Five variables were identified to determine the rate of adoption by individuals [16]. The five variables are perceived attributes of innovations, type of innovation-decision, communication channel, nature of the social system, and the change agent's promotion efforts; see Fig. 7.

The perceived attributes of innovation (relative advantage, compatibility, complexity, trialability, and observability) mainly influence the persuasion stage discussed above. Notably, most of these attributes cause negative perceptions for the three participants who reject to adopt Blackboard system. They perceived the system not compatible with their courses, not useful to them, complex to deal with. These are perception which is not really reflecting the actual system. To prove these perceptions reflects the actual system or not, more samples and examinations are needed. In our case we have two instructors adopt the system successfully compared to unsuccessful adoption by the three instructors. In the case of the students (the sample of 923), this cannot be examined independently because the main problem is that most of their instructors do not provide learning material on the Blackboard system.

The second variable that determines the rate of an innovation adoption is the type of the decision made. In our case at KAU, the use of Blackboard in the teaching process is optional. Notably, there are participants, in our study, who requested not to leave the decision of using the Blackboard optional despite they do not use it! They want it to be mandatory for all! However, to what extent will they be successful in their adoption if the decision is made mandatory?

An interesting question may be raised and urges answering by the decision makers at KAU. What about if the decision is made mandatory to use Blackboard by all instructors, for what purposes will they use it? This leads to draw the attention to the strategy of the organization. Have the organization (KAU Deanship of E-learning and faculties) defined well established, clear and targeted strategies? It is very important to refer to the DOI five stages of innovation process in an organization, see Fig. 1. Various concepts need to be well defined at the organizational level to understand how to proceed. What about the leader characteristics and attitude toward change, characteristics of the organizational structure (centralization, complexity, formalizations, interconnectedness, organizational slack, size), and the external characteristics of the organization/system openness. These are variables related to organizational innovativeness [16]. We will not go further discussing the issue at the organizational level because it is out of the scope of the current paper. The issue will be investigated and discussed in future work. However, in order to show an example of important issues for the decision makers to take into consideration when planning to make the use of Blackboard mandatory. What are the purposes of using Blackboard?

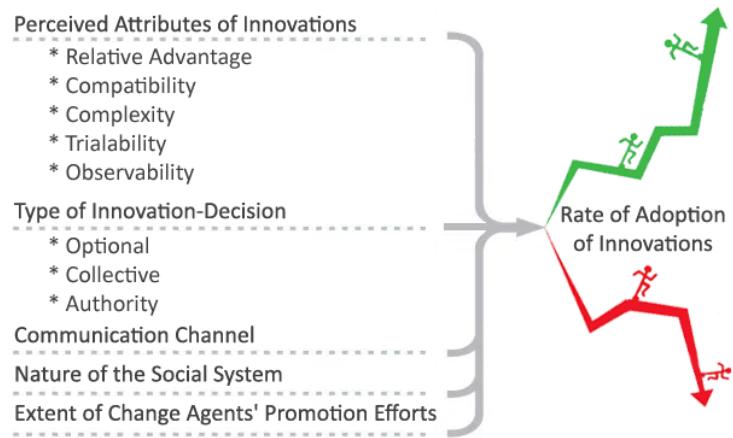

**Fig. 7. Variables Determining the Rate of Adoption of Innovation.**

Blackboard is a computer technology. It's made mainly to serve online education. However, how this technology can be used in traditional face-to-face (F2F) education? Active learning, flipped classrooms, blended learning, and similar other terms exist in relation to the use of technology to serve education. Which type of education can be used? Will the decision makers direct the instructors to use it as an additional tool without making any reduction of classes? If so, this is called "technology-enhanced" education. This type of education is perceived as additional effort and extra timing by the current study participants who reject adopting the Blackboard system! The second case, we may reduce the F2F contact time to be replaced with extra time online and this is called "blended learning". This type of education is what we personally prefer and practice with our students. Suppose a course takes two classes, each class is two hours duration, and the total is 4 hours a week. Consider meeting face to face takes 2 hours and the other 2 hours to spend online using Blackboard. An instructor in this case and the students as well will be encouraged (and to some extent enforced) to use the Blackboard.

However, to what extent, for example, a dean who strictly cannot tolerate skipping F2F classes will be able to accept this type of education? What about if an instructor makes the F2F classes optional and the main classes are online? See figure 8. More or less of these situations should be raised to consider the whole image of the adoption process at the organizational level.

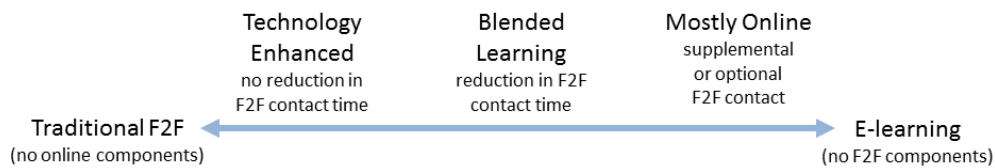

**Fig. 8. Spectrum of course-delivery modalities in higher education** [3].

Interestingly, one case in our study can be demonstrated as an example of change agent's promotion efforts. It is the case with the participant (Fahad) who successfully adopts the Blackboard system and played a significant role in influencing the second participant (Adel) to adopt the system. This case with other similar cases can be used as an internal change agent within the KAU. The efforts made by similar case positively influence the other users to do the same because it shows real cases like theirs. Recently, E-Learning Deanship at KAU has organized an award for excellence and creativity in using the Blackboard. Successful cases in adopting and using the Blackboard are awarded a prize S.A.R. 10,000 each [25]. This is good to present these successful cases to motivate others. However, more efforts are needed to group all these cases to represent a local change agent for the purpose to accelerate the process of the Blackboard adoption and use in each faculty.

### d. The 'FORCE' strategy

Based on the current study outcomes and our personal experience in teaching the computer graphics course (CPIT-285), we have developed the FORCE strategy. The FORCE strategy is an instructor-student direction to use Blackboard. It is not meant the literal meaning of the force but the way it is used does not make the use of Blackboard optional for the students. The word 'FORCE' stands for *Focus, Organize, Reduce, Communicate, and Enrich with graphics*, see Fig. 8. This strategy is to be used by an instructor dealing with students on Blackboard as follows.

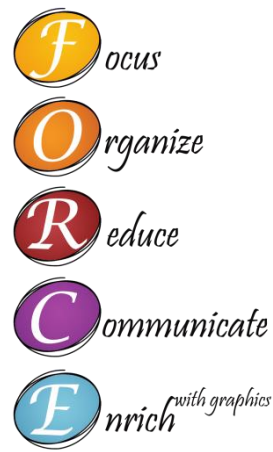

**Fig. 9. The FORCE Strategy.**

Focus on one place, the course page on Blackboard. Do not distract your students using other systems or websites. Blackboard is rich with tools that can assist providing valuable course content. Limit using links that refer to other websites, YouTube videos etc. For example, HTML and iframe tools can be used to integrate other websites contents, display videos, etc., on the same pages of the course. Keep referring your students to the Blackboard; do not give them direct answers to their questions if the content on Blackboard provides answers. If an individual contact asking a question that requires explanation, direct him/her to the discussion board to post the question. Posting a question to the discussion board make you and your students concentrated on one place. If another student has the same question, it is available there on the discussion board, and so on.

Organize the course content on Blackboard. The main page should provide a map for the course content. Use content items, folders, modules to organize and structure the information. Do not upload (throwing) files wherever they go! The well organized and nice looking of the course content attract students to get involved.

Reduce the number of pages and menu items. Unfortunately, the default template of the Blackboard at KAU is messy. Students get bored if they cannot navigate easily. Reducing the pages and menu items never means the content is less. You can provide rich content with good organization and well presentation as explained earlier.

Communicate using the communication tools on Blackboard. On Blackboard there are announcements, course messages, discussion board, blogs and comments. Users can opt to be notified as well on their e-mails. Concentrate on using these communication tools to keep the students engaged with the Blackboard. As discussed earlier in the '*Focus*' item, do not distract your students using other communication tools such as WhatsApp, MyKAU App, etc. keep the students engaged with the Blackboard and feel that they are connected 24/7.

Enrich the content with graphics. The use of graphics enhances the course content and makes it easy to navigate. It gives nice looking and well presented with the appropriate use of the pictures. Use headers, icons, and display textual forms as images, see Fig. 10. You can use the teaching style tool on Blackboard to customize the course templet, activate the use of icons, and upload a graphical header. On the content pages, for example, provide for each lecture note a graphical entry. It makes the lecture notes content looking nice, attractive and easy to navigate.

## 5. Limitations

The current paper has limitations to be acknowledged. The study is limited in its sampling selection. The sample was selected only from the KAU foundation-year. The reason for targeting the foundation-year is because all the students are at the same level; no other factors can influence their usages such as field of study. Of course, students at computer faculty differ in dealing with computers from students at any other faculties. However, we acknowledge that this is a limitation of the current study and it will be considered for future investigation with taking the faculties differences into account. In addition, the sampling did not cover female students and instructors. The results might differ in female section compared to male section. Therefore, the gender factor needs to be considered for future investigation.

**Fig. 10.   Screenshot of the Computer Graphics Course (CPIT-285) on Blackboard.**

## 6. Conclusions and Future Work

The current paper is a part of a diffusion research. It discussed the adoption process of the LMS at KAU at the individual level (instructors and students). Based on a quantitative investigation surveying about one thousand students followed by qualitative investigation with five instructors, the process of LMS adoption is not mature. Various factors involved in this process. Studying such a phenomenon in an organization is useless without considering the organizational factors. In this paper, the importance of studying the LMS adoption at the organizational level is demonstrated. Further investigation at the organizational level based on the current findings will be carried out and reported in future work.

# تقييم استخدامات نظام إدارة التعلم الإلكتروني في جامعة الملك عبد العزيز واقتراح استراتيجية القوة "FORCE" لدعم الاستخدام

رائد عبدالله الغامدي[1] و عادل باحداد[2]

[1] قسم تقنية المعلومات، و[2] قسم نظم المعلومات، كلية الحاسبات وتقنية المعلومات، جامعة الملك عبدالعزيز، جدة، المملكة العربية السعودية

raalghamdi8@kau.edu.sa

*المستخلص*. منذ بداية العام الدراسي ١٤٣٥هـ (سبتمبر ٢٠١٤م)، تم تطبيق نظام إدارة التعلم الإلكتروني "Blackboard" وإتاحته لجميع طلاب وطالبات جامعة الملك عبد العزيز، وبعد ثلاث سنوات من التطبيق يتم إجراء هذه الدراسة لتقييم الاستخدامات الحالية لهذا النظام في جامعة الملك عبد العزيز. تم جمع بيانات من طلاب السنة التحضيرية، من ٩٢٣ طالبًا، والذي يمثل حوالي ثلث العدد الإجمالي لشطر الطلاب للعام الدراسي ٢٠١٦/٢٠١٧م. استنادًا إلى الأدلة الإحصائية من ردود الطلاب على أسئلة الاستطلاع، فإن ٧٨٪ من الطلاب غير نشطين في استخدام نظام إدارة التعلم الإلكتروني بلاك بورد؛ ولذا اُتبع في هذا الاستطلاع إجراء مقابلات مع خمسة من أعضاء هيئة التدريس للسنة التحضيرية من أجل الحصول على تفسيرات لنتائج استطلاع الطلاب. تشير نتائج الدراسة إلى أهمية إحداث تغييرات على المستويين الفردي والمؤسسي من أجل تحسين الوضع القائم، ولهذا تم الاعتماد على نظرية نشر الابتكار (DOI) كقاعدة نظرية لهذه الدراسة، لأنها تقدم تفسيرات على المستويين الفردي والمؤسسي. وختامًا وبالاستناد على نتائج الدراسة الحالية وخبرة المؤلف في تدريس مقرر الرسومات بالحاسوب باستخدام "Blackboard"، تم اقتراح استراتيجية تسمى: قوة "FORCE"؛ لتسريع وتيرة تبني استخدام "Blackboard".